\documentclass[11pt,a4paper]{article}
\usepackage{jcappub}
\pdfoutput=1
\usepackage[english]{babel}
\usepackage{amsmath,amssymb,amsfonts,bm,bbm,slashed, subdepth}
\usepackage{graphicx}
\usepackage[sort&compress]{natbib}
\usepackage{xcolor}
\usepackage[normalem]{ulem}
\usepackage{times}
\usepackage{hyperref}
\usepackage{cleveref}
\definecolor{red}{rgb}{1.0, 0, 0}
\usepackage{hyperref}
\usepackage{enumerate}
\usepackage{epsfig, subfigure}
\usepackage{setspace}
\usepackage{booktabs, tabularx}
\usepackage{units}
\usepackage{hyperref}

\newcommand\ion[2]{#1{\scshape{#2}}}

\newcommand{\be}{\begin{equation}}
\newcommand{\ee}{\end{equation}}
\newcommand{\ba}{\begin{array}}
\newcommand{\ea}{\end{array}}
\newcommand{\bea}{\begin{eqnarray}}
\newcommand{\eea}{\end{eqnarray}}
\newcommand{\balg}{\begin{align}}
\newcommand{\ealg}{\end{align}}
\newcommand{\bit}{\begin{itemize}}
\newcommand{\eit}{\end{itemize}}
\newcommand{\trm}[1]{\textrm{#1}}

\newcommand{\cmb}{\hbox{\sc cmb}}
\newcommand{\edges}{{\sc edges}}
\newcommand{\wdm}{\hbox{\sc wdm}}
\newcommand{\ti}{\hbox{\sc ti}}
\newcommand{\cdm}{\hbox{\sc cdm}}
\newcommand{\dm}{\hbox{\sc dm}}
\newcommand{\dmf}{\hbox{\sc dmf}}
\newcommand{\Mps}{\hbox{$M_{\rm ST}$}}
\newcommand{\lya}{\hbox{Lyman-$\alpha$}}
\usepackage{enumitem}

\newcommand{\Mpc}{\trm{\Mpc}}
\newcommand{\yr}{\trm{\yr}}
\newcommand{\eV}{\trm{\eV}}
\newcommand{\Ts}{\hbox{$T_\mathrm{s}$}}
\newcommand{\Tk}{\hbox{$T_\mathrm{k}$}}

\title{\boldmath Constraining structure formation using {\sc edges}}

\author[a,b]{Matteo Leo,}
\author[b]{Tom Theuns,}
\author[b]{Carlton M. Baugh,}
\author[b]{Baojiu Li}
\author[a]{and Silvia Pascoli}

\affiliation[a]{
 Institute for Particle Physics Phenomenology, Department of Physics, Durham University, Durham DH1 3LE, U.K.}
\affiliation[b]{
 Institute for Computational Cosmology, Department of Physics, Durham University, Durham DH1 3LE, U.K.}
 
\emailAdd{matteo.leo@durham.ac.uk}

\abstract{
The {\em experiment to detect the global epoch of reionization signature} (\edges) collaboration reported the detection of a line at 78~MHz in the sky-averaged spectrum due to neutral hydrogen (\ion{H}{i}) 21-cm hyperfine absorption of cosmic microwave background  (\cmb) photons at redshift $z\sim 17$. This requires that the spin temperature of \ion{H}{i} be coupled to the kinetic temperature of the gas at this redshift through the scattering of \lya\ photons emitted by massive stars. To explain the experimental result, star formation needs to be sufficiently  efficient at $z\sim 17$ and this can be used to constrain models in which small-scale structure formation is suppressed (\dmf\ models), either due to dark matter free-streaming or non-standard inflationary dynamics. We combine simulations of structure formation with a simple recipe for star formation to investigate whether these models emit enough Lyman-$\alpha$ photons to reproduce the experimental signal for reasonable values of the star formation efficiency, $f_\star$. We find that a thermal warm dark matter (\wdm) model with mass $m_\mathrm{WDM}\sim 4.3\,\mathrm{keV}$ is consistent with the timing of the signal for $f_\star \lesssim 2\%$. The exponential growth of structure around $z\sim 17$ in such a model naturally generates a sharp onset of the absorption. A warmer model with $m_\mathrm{WDM}\sim3\,\mathrm{keV}$ requires a higher star formation efficiency, $f_\star\sim 6\%$, which is a factor of few above predictions of current star formation models and observations of satellites in the Milky Way. However, uncertainties in the process of star formation at these redshifts do not  allow to derive strong constrains on such models using 21-cm absorption line. The onset of the 21-cm absorption is generally faster in \dmf\ models compared to cold dark matter (\cdm) models, unless some process significantly suppresses star formation in halos with {masses below $\sim 10^8\,h^{-1}\,\mathrm{M}_\odot$}.
}

\begin{document}
\hfill{IPPP/19/72}\\
\maketitle
\flushbottom
\section{\label{sec:level1}Introduction}
Cosmic gas between us and the surface of last scattering can produce a global, redshifted \lq 21-cm\rq\ line originating from the hyperfine transition of neutral hydrogen (\ion{H}{i}). This line appears in emission or absorption in the spectrum of the cosmic microwave background (\cmb), depending on whether the spin temperature of the gas, \Ts\  (see Section~\ref{sec:21-cm_theory} for a definition of this quantity), is larger or smaller than the temperature of the \cmb\ photons, $T_\gamma$. Several processes conspire to make \Ts\ deviate from $T_\gamma$ following recombination of the Universe at redshift $z\sim 1100$. Initially, Compton heating of electrons left over after recombination keeps the kinetic temperature of the gas, \Tk, coupled to the temperature of the \cmb, $\Tk\sim T_\gamma$. Eventually, \Tk\ decouples from $T_\gamma$ below $z\sim 300$, and the gas temperature falls adiabatically as the Universe expands, $\Tk\propto (1+z)^2$, whereas $T_\gamma\propto (1+z)$ \cite{Peebles1968}. Collisions between neutral hydrogen atoms keep $\Ts\sim \Tk$ so that $\Ts<T_{\gamma}$, and the intervening gas appears in 21-cm absorption against the \cmb. Below $z\sim 30$, the \ion{H}{i} collision rate becomes too low to keep \Ts\ coupled to \Tk, the spin temperature increases to $T_\gamma$, and the gas becomes transparent to 21-cm photons. As the first sources of \lya\ photons - such as {\em e.g.} massive stars - appear around $z\sim 20$, scattering of \lya\ photons off \ion{H}{i} atoms, again couple \Ts\ to \Tk\ through the Wouthuysen-Field (hereafter WF) effect \cite{Wouthuysen1952,Field1958}. This results in $\Ts\sim\Tk$, and since
$\Tk<T_{\gamma}$, the cosmic gas once more appears in absorption against the \cmb. The absorption signal weakens and briefly turns into emission due to heating by X-rays emitted by early black holes and/or X-ray binaries \cite{Tozzi2000,Chen2004,Ricotti2005}. It is finally wiped out following reionization of the \ion{H}{i}. For a more in-depth discussion and original references, see {\em e.g.} \cite{Barkana2001,Furlanetto2006, Furlanetto2006_b, Pritchard2012, Lewis2007, Burns2017}.

The \edges\footnote{Experiment to Detect the Global Epoch of Reionization  Signature. \\\url{https://www.haystack.mit.edu/ast/arrays/Edges/index.html}.} collaboration has reported \cite{Bowman2018} the detection of an absorption line centred at 78~MHz in the sky-averaged spectrum, which they interpret as being due to \ion{H}{i} 21-cm absorption at $z\sim 17$ against the \cmb\, with \Ts\ coupled to \Tk\ by the WF-effect. The depth of the detected absorption line corresponds to an \lq antenna temperature\rq\ difference of $\delta T^{\mathrm{min}}_\mathrm{b} \sim -500\,\mathrm{mK}$, and the onset of the absorption has $\delta z\sim 3$, where $\delta z$ is the redshift width from $\delta T_\mathrm{b} = 0$ to $\delta T_\mathrm{b} = \delta T^{\mathrm{min}}_\mathrm{b}$. The observed line is stronger than expected by a factor of $\sim 2$. The line strength is in principle simply set by the ratio between \Tk\ and $T_\gamma$, which are both well known 
in the context of the standard cosmological model ($\Lambda$\cdm). The unexpected observed value may signal the need for new physical mechanisms \cite{Barkana2018} that produce an enhancement in the value of the ratio $T_\gamma$/\Tk\ at $z\sim 17$ respect to that expected from $\Lambda$\cdm\ calculations, for example non-gravitational dark matter (\dm)-baryon interactions or the presence of extra sources of radio emission (see e.g. \cite{Feng2018, Barkana2018_b,Fraser2018, Berlin2018,Munoz2018,Slatyer2018, Sharma2018, Sikivie2018, Houston2018, Li2019, Jia2019, Chatterjee2019,Mishra2019, Bhatt2019} for an incomplete list of references on these topics). Recently, it has been pointed out that polarized foreground contamination may produce an enhanced 21-cm absorption line \cite{Spinelli2019}.
More worryingly, \cite{Hills2018} suggests that the shape - and even the reality - of the signal is potentially strongly affected by how foregrounds were modelled by \cite{Bowman2018} (but see the reply by \cite{Bowman2018b}). The re-analysis of the data by \cite{Singh2019} results in a weaker
absorption signal, but the onset of the absorption remains relatively sudden. While acknowledging these concerns, the {\sc edges} signal has been used to constrain a wide range of non-standard cosmological scenarios, see {\em e.g.} \cite{DAmico2018,Hill2018,Hektor2018, Mitridate2018,Wang2018,Li2019_b,Yang2019}.

The shape of the downturn of the line is a measure of the rate at which stars build up a background of \lya\ photons. In a \cdm\ universe, 
the first stars form in \dm\ halos with virial mass $M_h\sim 10^6\, h^{-1}$~M$_\odot$, when \ion{H}{i} forms H$_2$ which allows the gas to cool and become self-gravitating \cite{Abel2002}. Such \lq population three\rq\ (Pop.~III) stars are thought to form one -- or at most a few -- per halo, and are generically expected to be more massive than the typical star formed today because the Jeans mass in the hotter star forming gas is higher than today \cite{Susa2014,Hirano2015,Stacy2016}. Such massive stars are hot and hence radiate copious \lya\ photons \cite{Schaerer2002}. As these stars enrich their surroundings with metals that help cool gas and promote H$_2$ formation, and as progressively more massive halos form, star formation is thought to become more similar to what it is today, with lower-mass \lq population two\rq\ (Pop.~II) stars forming in gas that initially cools atomically. This standard picture of the onset of star formation in the Universe results in a relatively gentle build-up of a background of \lya\ photons, resulting in a more extended onset of the 21-cm line than observed. Reconciling the \cdm\ model with the \edges\ signal therefore requires that only halos with $M_h\gtrsim 10^8h^{-1}$~M$_\odot$ contribute significantly to star formation, for example because star formation
in lower-mass halos is strongly suppressed due to energy injected by supernovae \cite{Mirocha2019, Kaurov2018}. 

An alternative way of making the onset of 21-cm absorption more rapid is to suppress matter fluctuations at small scales such that these lower mass halos simply do not form, by changing either the nature of the \dm\ or the physics of the very-early universe. If the \dm\ has a large free-streaming length, it smooths out small-scale structure below some characteristic damping scale $\lambda_d$, because of the intrinsic velocities of the \dm\ particles \cite{Bode2000, Colin2000, Hansen2002, Viel2005, Dodelson1994, Dolgov2002, Asaka2007, Enqvist1990, Shi1999, Abazajian2001, Kusenko2006, Petraki2008, Merle2015, Konig2016}. Such models are generically termed \lq warm dark matter\rq\ (\wdm) models\footnote{In the context of alternative \dm\ scenarios, a suppression of the gravitational clustering on small scales can be also achieved allowing \dm\ particles to have non-vanishing interactions (either with themselves \cite{Spergel2000} or with neutrinos/photons \cite{Boehm2005, Boehm2014}) or considering models where \dm\ is a scalar field with a macroscopic wave-like behaviour  \cite{Marsh2016, Veltmaat2016, Veltmaat2018}.}. Small-scale power can also be suppressed due to non-standard inflationary dynamics \cite{Kamionkowski2000,Hong2015,Hong2017, Leo2018, Enqvist2019}. We will refer to a model in which power is significantly suppressed below some scale $\lambda_d$ (compared to \cdm) generically as a model with damped matter fluctuations (\dmf), and the co-moving mass in a volume with radius $\lambda_d$ as the \lq damping mass\rq, $M_d$. The onset of {\em star} formation may be very different in \dmf\ models, because the first structures to collapse are extended filaments with a mass of order of the damping mass, rather than halos \cite{Gao2007} (see also \cite{Hirano2017}). The very different nature of the \dm\ potential wells in which the first stars form is likely to affect the nature of these stars - for example their mass - as well as the abundance of such stars - {\em i.e.} the total number of stars formed per unit volume. Making accurate {\em quantitative} predictions for how this affects the 21-cm signal is challenging. However, generically we expect these stars to form more abundantly and be of higher mass compared to \cdm\ models, mainly because the filaments can collect a large amount of gas {\em before} any stellar processes can limit gas accretion. The latter is because there is no stellar feedback in progenitors as a result of the progenitor halos themselves not forming (see \cite{Gao2007} for more details along these lines). We therefore expect any 21-cm signal to build up rapidly.

Independent motivation for examining \dmf\ models comes from particle physics. For example, sterile neutrinos act as \wdm\ and have been proposed to explain the observed baryon asymmetry of the Universe (\cite{Shaposhnikov2007}, see \cite{Boyarsky2018} for a recent review). In addition, \wdm\ has been proposed as a solution to some perceived astrophysical problems related to the number density and concentration of dwarf galaxies \cite{Bode2000,Colin2000,Hansen2002} (see \cite{Weinberg2015} for a recent review). Constraints on the \lq warmness\rq, {\em i.e.} the scale $\lambda_d$ below which \wdm\ suppresses structure, is often quoted in terms of the mass, $m_{\wdm}$, of the thermal \wdm\ particle with the same value of $\lambda_d$. 
Observations of the \lya\ forest \cite{Garzilli2015, Viel2005, Garzilli2019,Irsic2017,Murgia2017} and constraints resulting from the observed satellite luminosity function of the Milky Way galaxy \cite{Lovell2014,Lovell2017} robustly exclude thermal \wdm\ candidates with masses lower than $m_\mathrm{WDM}\sim 2\,\mathrm{keV}$. Indeed, in these models {\em less} structure forms than observed. 

The timing of the \edges\ signal constrains $\lambda_d$: too much suppression delays structure formation and hence
the \lya\ background also builds up too late \cite{Schneider2018, Lidz2018, Safarzadeh2018, Chatterjee2019,Boyarsky2019, Lopez-Honorez2019}. Here we re-examine this constraint.
This paper is structured as follows. We begin with a brief overview of how the emission of Lyman-$\alpha$ from young galaxies is related
to the 21-cm signal through the WF-effect in Section~\ref{sec:21-cm_theory}. The \dm\ models that we use are introduced in Section~\ref{sec:models_simulations}, together with details of the numerical simulations for calculating the rate of formation of \dm\ structures in which the young galaxies form. The 21-cm signal corresponding to the different models is discussed in Section~\ref{sec:starformLyalpha}. Finally, Section~\ref{sec:summary} summarizes our findings.

\section{Modeling the 21-cm signal}
\subsection{Hyperfine 21-cm absorption against the \cmb}
\label{sec:21-cm_theory}
This section briefly reviews the well-known physics behind 21-cm \ion{H}{i} hyperfine absorption against the \cmb\, see {\em e.g.} \cite{Pritchard2012}. The strength of the absorption depends on three temperatures, ({\em i}) the spin temperature, \Ts,  ({\em ii}) the kinetic temperature of the gas,  \Tk, and ({\em iii}) the \cmb\ temperature, $T_\gamma$. When \ion{H}{i}
atoms are in the electronic ground state, \Ts\ sets the fraction of atoms that are in the higher energy triplet state (proton and electron have parallel spin, state $n_1$) compared to the singlet state (anti-parallel spins, state $n_0$),
\begin{equation}
 \frac{n_1}{n_0} =  \frac{g_1}{g_0} \,\exp{\left(-\frac{T_\star}{\Ts}\right)}\,.
\end{equation}
Here, $T_\star$ is the atomic constant $T_\star \equiv hc/(k_\mathrm{B}\,\lambda_{21}) \approx 0.068\,\mathrm{K}$, with $h$ Planck's constant, $k_\mathrm{B}$ Boltzmann's constant, $c$ the speed of light, and $\lambda_{21}\approx 21.1~{\rm cm}$ the wavelength of the 21-cm line; $g_1/g_0=3$ is the ratio of degeneracy levels of the triplet to the singlet state. In equilibrium, \Ts~$=~$\Tk~$=T_\gamma$, and neutral gas absorbs 21-cm photons from the \cmb\ at the same rate that it emits such photons making the gas transparent. When $\Ts<T_\gamma$, more photons are absorbed than emitted, and intervening gas appears in absorption against the \cmb. The intensity of the absorption signal strength depends on \Ts.

It is customary in radio astronomy to quantify the specific intensity of a signal at frequency $\nu$, $I_\nu$, in terms of its \lq apparent brightness\rq\, or \lq antenna temperature\rq. This is the temperature of a black body that has the same value of $I_\nu$ in the Rayleigh-Jeans part of the spectrum, $I_\nu=2k_\mathrm{B}T\nu^2/c^2$. The strength of the 21-cm absorption is then the temperature difference, $\delta T_\mathrm{b}$, between the brightness temperature of the signal and that of the \cmb. It is related to \Ts\ by (as given in \cite{Pritchard2012}),
\begin{equation}
\delta T_\mathrm{b} \approx 27 \,\mathrm{mK}\,x_\mathrm{HI}(z) \,  \,\left(\frac{\Omega_{b}h^2}{0.023}\right)\,\left(\frac{0.15}{\Omega_{m} h^2}\,\frac{1+z}{10}\right)^{1/2}\left(1-\frac{T_\gamma(z)}{T_\mathrm{s}(z)}\right)\,.
\label{eq:dtb}
\end{equation}
Here, $x_\mathrm{HI}(z)\approx 1$ at $z\sim 17$ is the fraction of gas in the form of \ion{H}{i} (see {\em e.g.} \cite{Furlanetto2006}),  $T_\gamma = T^0_\gamma\, (1+z)$ is the \cmb\ temperature in terms of its value $T^0_\gamma\approx 2.73~{\rm K}$ today; $\Omega_b$ and $\Omega_m$ are the cosmological baryon density and total matter density
in units of the critical density, respectively. $\delta T_\mathrm{b}<0$ occurs for $\Ts<T_\gamma$, which signals absorption.

The situation where $\Ts<T_\gamma$ arises when \Ts\ gets coupled to \Tk, because as the Universe cools adiabatically, \Tk\ drops faster than $T_\gamma$, so that $\Ts\approx \Tk$ results in $\Ts<T_\gamma$.
Such coupling can be caused by collisions in sufficiently dense regions and by scattering of \lya\ photons produced by early sources such as hot stars through the WF-effect. The basic physics behind the WF-effect is that when an \ion{H}{i} atom in the $n=1$ electronic ground state absorbs and then re-emits a \lya\ photon, it can flip from the singlet to the triplet state or vice-versa. However, when $\Ts=\Tk$, then there should be no net energy transfer between the hyperfine states and the gas, therefore \lya\ scattering will couple \Ts\ to \Tk. The coupling strength depends on
atomic constants and the specific mean intensity $J_\alpha$ of the radiation at the \lya\ wavelength ({\em e.g.} \cite{Pritchard2006,Hirata2006,Barkana2005}),
\begin{eqnarray}
1-\frac{T_\gamma}{\Ts} &=& \frac{x_\alpha}{1+x_\alpha} \,\left(1-\frac{T_\gamma}{\Tk}\right)\nonumber\\
\label{eq:x_tot}
x_\alpha &=&  \frac{16\pi^2 T_\star e^2 f_\alpha}{27 A_{10} T_\gamma m_e c}\, S_\alpha J_\alpha\,.
\label{eq:xa}
\end{eqnarray}
Here, $f_\alpha = 0.4162$ is the oscillator strength of the Lyman-$\alpha$ line, $A_{10} = 2.85\times10^{-15}\,\mathrm{s}^{-1}$ is the Einstein coefficient of the 21-cm transition, $e$ and $m_e$ are the electron charge and mass, respectively; $S_\alpha$ is a correction factor that accounts for spectral distortions for which we take $S_\alpha \simeq 1$ following \cite{Hirata2006}. The required Lyman-alpha flux needed for effective WF coupling was estimated {\em e.g.} in \cite{Ciardi2003}.

If the \ion{H}{i} atom had only two electronic energy levels, then $J_\alpha$ would simply be the background of \lya\ photons produced by early sources. However, \lya\ photons can be produced by the absorption of photons in the higher Lyman series, followed by a radiative cascade. We also need to account for photons redshifting out of, and into, the \lya\ transition. Taking all of this into account
relates $J_\alpha$ to the emissivity of the sources, $\epsilon_\nu(z)$, as 
\cite{Pritchard2006,Hirata2006,Barkana2005}
\begin{equation}
J_\alpha(z) = \frac{c\,(1+z)^2}{4\pi}\,\sum^{23}_{n=2} f_n \, \int^{z_{\mathrm{max,}\,n}}_z dz^\prime\, \frac{\epsilon_{\nu}(z^\prime)}{H(z^\prime)}.
\label{eq:Ja}
\end{equation}
Here, the $f_n$ are atomic constants related to the radiative cascade (see {\em e.g}.~\cite{Hirata2006} for the numerical values of $f_n$), $H(z)$ is the Hubble constant, and $z_{\mathrm{max,}\,n}$ is given by \cite{Pritchard2012} as
\begin{equation}
z_{\mathrm{max,}\,n} = (1+z)\, \left(\frac{1-(1+n)^{-2}}{1-n^{-2}}\right) -1\,.
\end{equation}
We will assume that the sources of UV-photons are hot stars that form in collapsed structures. Therefore to compute $\epsilon_\nu(z)$, we first need to know the fraction of mass that collapses into bound structures
in which star formation can proceed, $f_{\rm coll}(z)$. This fraction depends on cosmology and on the shape of the power spectrum, as we examine next.

\subsection{Structure formation}
\label{sec:models_simulations}
We want to contrast the expected 21-cm signal in \cdm\ models to that in alternative models in which the power below some co-moving damping scale $\lambda_d$ is suppressed compared to \cdm. 
We begin by describing how we calculate $f_{\rm coll}(z)$ in \cdm\ models.
\subsubsection{\cdm\ models}
To compute $f_{\rm coll}$, we start by computing the evolution of the halo mass function, $n(M,z)$,
for which we use the Sheth-Tormen (ST) extension \cite{Sheth1999} of the Press-Schechter (PS) formalism \cite{Press1974,Bond:1991,Zentner:2007,Benson2013}. The halo mass function is the (co-moving) number density of halos of mass $M$ at redshift $z$, and is given by
\begin{equation}
\frac{d n_{\cdm}}{d \ln (M)} = \frac{1}{2} \,\frac{\bar{\rho}^0_{m}}{M} \, {f(\nu)} \frac{d \ln(\nu)} {d \ln(M)}\,,
\label{eq:PShalomass}
\end{equation}
where $\bar{\rho}^0_{m}$ is the mean co-moving matter density and 
\begin{equation}
\nu = \frac{\delta^{2}_{c,0}}{\sigma^2(R) D^2(z)}\,.
\end{equation}
Here, $\delta_{c,0} = 1.686$, $D(z)$ is the linear growth factor normalized to $D=1$ at $z=0$, and $\sigma^2(R)$ is the mass-variance on scale $R$,
\begin{equation}
\sigma^2(R) = \int  \frac{d^3\mathbf{k}}{(2\pi)^3} P_\mathrm{CDM}(k) \tilde{W}^2(k|R)\,.
\label{eq:variace}
\end{equation}
In this expression, $P_{\cdm}(k)$ is the linear matter power spectrum at $z=0$ and $\tilde{W}(k|R)$ is (the Fourier transform of) the filter function. We use a spherical top-hat (other window functions have been discussed in the literature, see {\em e.g.} \cite{Bond:1991,Benson2013,Schneider2013,Schneider2015, Leo2018_c}), given in real space by
\begin{equation}
W(r|R) = \begin{cases}
\frac{3}{4\pi R^3}\quad& \,\mathrm{if}\quad r \leq R\\
\,\,\,\,0 \quad& \,\mathrm{if}\quad r > R\\
\end{cases}. 
\end{equation} 

The ST formalism uses the ellipsoidal collapse model of \cite{Sheth1999} to compute $f(\nu)$. This function is well approximated by
\begin{equation}
f(\nu) = A \, \sqrt{\frac{2q\nu}{\pi}} \left( 1+ (q\nu)^{-p} \right) \exp({-q\nu/2}),
\end{equation}
with $A=0.3222$, $p=0.3$ and $q=0.707$.

The damping mass $M_d$ is effectively zero in \cdm\, and consequently all dark matter is in collapsed objects of some mass
at any $z$, $f_{\rm coll}\approx 1$. However, the numerous low-mass dark matter halos that form at high $z$ will not contribute significantly to star formation and hence are irrelevant for computing $\epsilon_\nu$. The reason is that, if the virial temperature, $T_{\rm vir}$, of a halo is too low, the gas is too cold to cool and form stars. For $T_{\rm vir}\sim 8000$~K, the gas is thought to be hot enough to cool via the formation of H$_2$ \cite{Abel2002}, once $T_{\rm vir}\sim 10^4$~K, gas can cool by atomic transitions in \ion{H}{i} (see {\em e.g.} \cite{Barkana2001} for more details). To account for this, we will only include \dm\ halos above a given minimum mass\footnote{A given minimum halo mass can be  converted to a corresponding minimum virial temperature of the star forming halo using Eq.~(26) in~\cite{Barkana2001}.} when computing the collapsed fraction $f_{\rm coll}$. Below we will illustrate \cdm\ results for $M_{\rm min}=10^7~h^{-1}{\rm M}_\odot$ and $M_{\rm min}=10^8~h^{-1}{\rm M}_\odot$, denoting these models by \lq \cdm-7\rq, and \lq \cdm-8\rq\, respectively. Given $n(M,z)$, we can compute the collapsed fraction for these models as
\begin{equation}
f_\mathrm{coll}(z) =\frac{1}{\bar{\rho}^0_{m}}\int^\infty_{M_\mathrm{min}} dM \,M\,\frac{d n_{\cdm}(z)}{d M}\,.
\end{equation}
The evolution of $f_\mathrm{coll}$ in the range of redshifts considered in our analysis is shown in Figure~\ref{fig:1b} for \cdm-7 and  \cdm-8. As expected, the values of the collapsed fraction are always larger in \cdm-7 than \cdm-8 because more halos are included in the calculation of the former. An  interesting difference between the two \cdm\ models is that the build-up of structure in \cdm-8 is more rapid than in \cdm-7. We will return on this aspect when discussing the results in Section~\ref{sec:starformLyalpha}.
\subsubsection{\dmf\ models}

\begin{figure}[t!]
\advance\leftskip-.15cm
\advance\rightskip-.5cm
\subfigure[][\dmf\ $P(k)$ (shown as ratios w.r.t. \cdm)]
{\includegraphics[width=.52\textwidth]{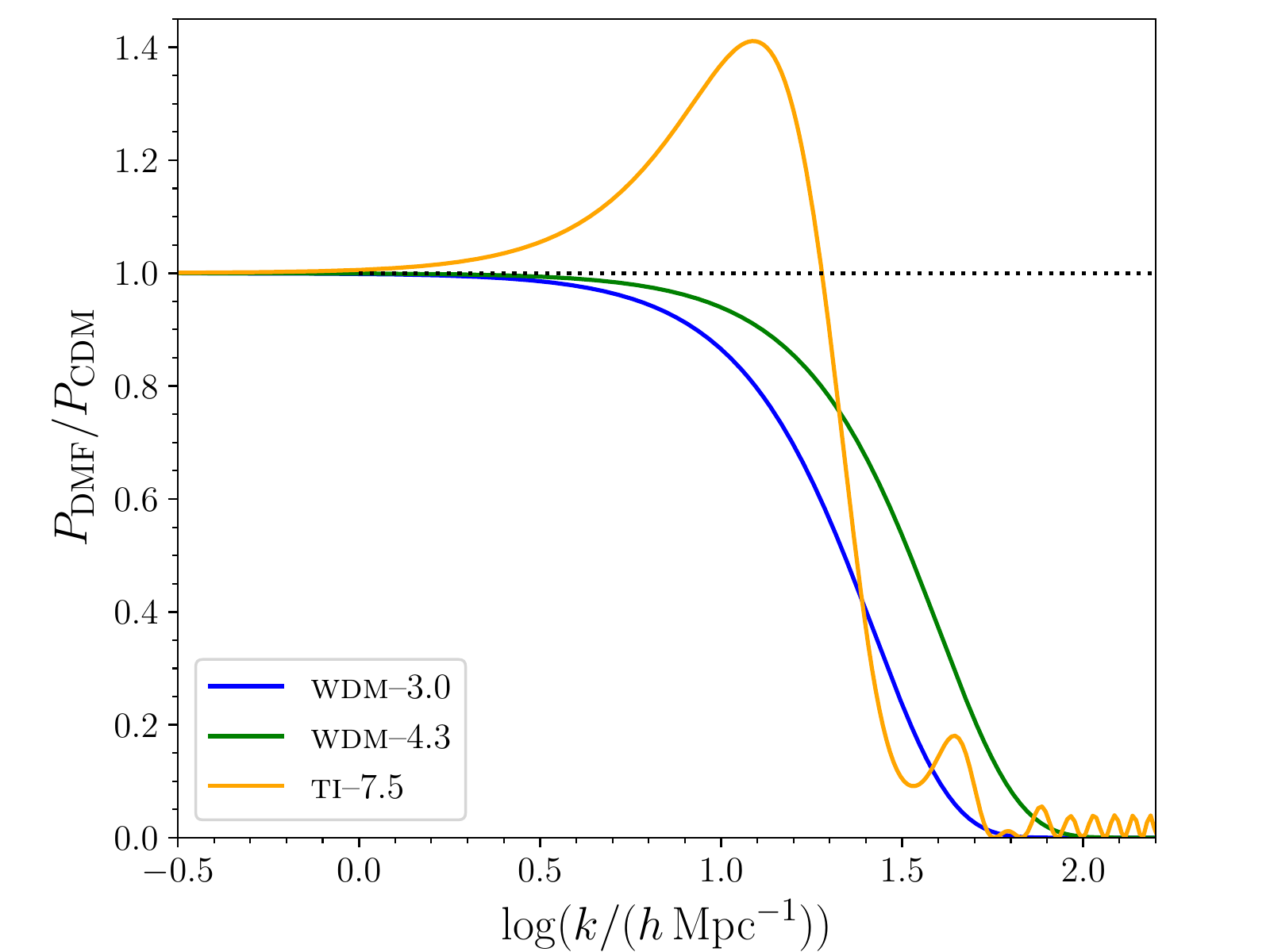}\label{fig:1a}}\hspace{-4.ex}
\subfigure[][\dmf\ collapsed fraction]{\includegraphics[width=.52\textwidth]{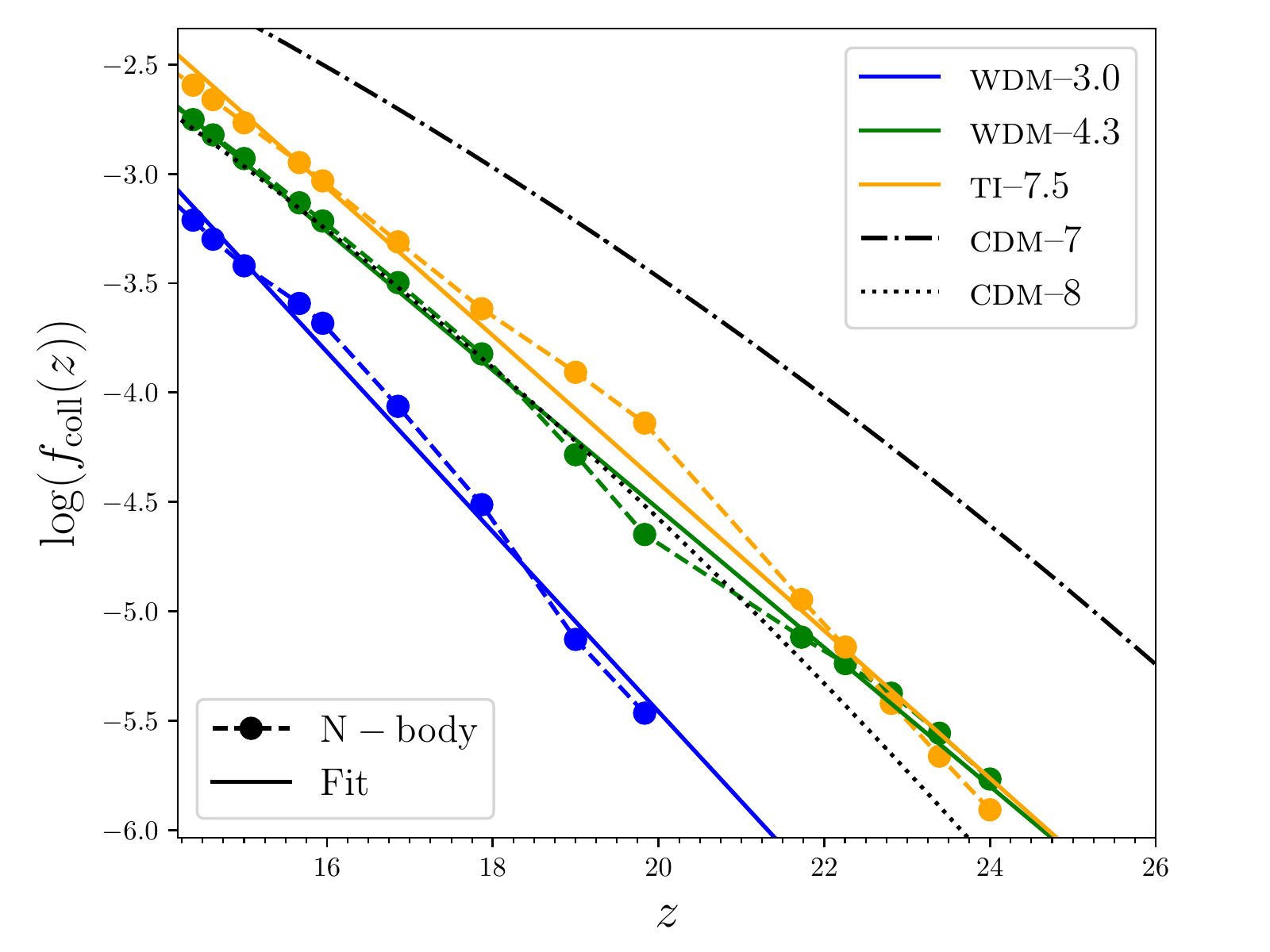}\label{fig:1b}}\\[-2. ex]
\caption{(a) Ratios w.r.t. \cdm\ of the linear theory power spectra for \wdm\ and \ti\ models (as labelled). (b) Evolution of the collapsed fraction, $f_\mathrm{coll}(z)$, for the three \dmf\ (colour) and the two \cdm\ (black) models  considered in this analysis. In the case of \dmf, symbols show the results from the simulation, while solid lines show those from the parametrisation Eq.~(\ref{eq:fit1}), with parameters reported in Table~\ref{table:best_fit}. The $f_\mathrm{coll}(z)$ for \cdm\ are obtained from the Sheth-Tormen extension of the Press-Schechter formalism (see main text for more details).}
\label{fig:1}

\end{figure}

\begin{table}
\centering
\begin{tabular}{|c|c|c|c|}
\hline
Model&  \wdm-3.0& \wdm-4.3& \ti-7.5\\
\hline\hline
$f_{14}$     & $1.02\times 10^{-3}$  & $ 2.34\times 10^{-3}$&$4.08\times 10^{-3}$\\
\hline
$\zeta$     & $0.95$  & $ 0.73$& $0.78$\\
\hline
\end{tabular}
\caption{Fitting parameters $f_{14}$ and $\zeta$ for the fit of Eq.~(\ref{eq:fit1}) to
the evolution of the fraction of mass in collapsed objects plotted in Fig.~\ref{fig:1b}, for the three \dmf\ models, \wdm-3.0, \wdm-4.3 and \ti-7.5.}
\label{table:best_fit}
\end{table}

For the \wdm\ models, we introduce an exponential cut-off in the power spectrum to mimic the effect of free-streaming, 
\begin{equation}
P_{\wdm}(k) = P_{\cdm}(k) \, \exp(-\lambda^2_d\,k^2)\,.
\label{eq:1}
\end{equation}
We examine two models, taking $\lambda_d = 0.038\, h^{-1}\,\mathrm{Mpc}$ and $\lambda_d = 0.025 \, h^{-1}\,\mathrm{Mpc}$, which correspond to two choices for the \wdm\ thermal-equivalent particle mass\footnote{More accurately, the two mass values are  $m_\mathrm{WDM}= 2.92\,\mathrm{keV}$ and $m_\mathrm{WDM}= 4.25\,\mathrm{keV}$.} $m_\mathrm{WDM}\sim 3\,\mathrm{keV}$ and $m_\mathrm{WDM} \sim 4.3\,\mathrm{keV}$. We will refer to these models as \wdm-3.0 and \wdm-4.3, respectively. Note that our Eq.~(\ref{eq:1}) is considered as an approximation of the real effect of the free-streaming on the linear matter power spectrum of \wdm\ models.  Power spectra that are more accurate than simply imposing an exponential cut-off can be generated using either Boltzmann codes such as {\sc class} \cite{Lesgourgues2011, Lesgourgues2011_b} or the transfer function proposed in  \cite{Bode2000, Viel2005}. The advantage of adopting an exponential cut-off resides in the fact that, the only free-parameter in the exponential ($\lambda_d$, see Eq.~(\ref{eq:1})) 
unequivocally identifies the scale of the damping. On the other hand, transfer functions as that in \cite{Bode2000, Viel2005}   depend on the particular \wdm\ model considered and are, in general, given in terms of particle physics parameters (such as the mass of the \wdm\ candidate), whose relation with the damping scale is more subtle than that displayed in Eq.~(\ref{eq:1}).    Nevertheless, we expect that our results on the 21-cm absorption signal will not change dramatically when considering  more accurate power spectra than those employed here. We additionally consider a thermal inflation (\ti) model with $k_\mathrm{b} = 7.5 \,\mathrm{Mpc}^{-1}$ ($k_\mathrm{b}$ represents the wave number above which the \ti\ linear power spectrum starts to deviate appreciably from that of standard \cdm, see \cite{Hong2015}), generated using the transfer function calculated by \cite{Hong2015, Hong2017},

\begin{equation}
\begin{split}
T_\mathrm{TI} (\xi) &= \cos\left[\xi\int^\infty_0 \frac{d\alpha}{\sqrt{\alpha(2+\alpha^3)}}\right]+6\xi\int^\infty_0 \frac{d\gamma}{\gamma^3}\int^\infty_0 d\beta \left(\frac{\beta}{2+\beta^3}\right)^{3/2} \sin\left[\xi\int^\infty_\gamma  \frac{d\alpha}{\sqrt{\alpha(2+\alpha^3)}}\right]\,,\end{split}
\end{equation}
where $\xi= k/k_b$ and the power spectrum is $P_\mathrm{TI}(k)=P_{\mathrm{CDM}}(k) T^2_\mathrm{TI}(k)$;
we will refer to this model as \ti-7.5. The suppression of power compared to \cdm\
is plotted in Figure~\ref{fig:1a} for these three models. For the \wdm\ models, we see that a larger value of $m_\mathrm{WDM}$ suppresses power on smaller scales ({\em green} versus {\em blue} curve). The power-spectrum of the \ti-7.5 model ({\em yellow} curve) is suppressed more strongly than \wdm-4.3 for  $k\gtrsim 25\,h\,\mathrm{Mpc}^{-1}$, however at wave numbers in the range $k\in[5,20]\,h\,\mathrm{Mpc}^{-1}$, the power in \ti-7.5 is {\em enhanced} compared to \cdm. This characteristic enhancement is one of the main features of this model, compared to \wdm. Its impact on the non-linear power spectrum and halo abundances has been studied by \cite{Leo2018}; the effects on structure formation of other models with two inflationary stages have also been investigated by \cite{Enqvist2019}.

Given these linear power spectra, we have performed \dm\ only cosmological simulations of structure formation, using the tree-PM N-body code Gadget-2 \cite{Springel2005}. Initial conditions were generated at $z=199$, an epoch in which all the wave numbers probed in the simulation are well inside the linear regime, 
using second-order Lagrangian perturbation theory with {\sc 2lpt}ic \cite{Crocce2006}. We choose a box of co-moving length $L_\mathrm{box}=5\,h^{-1}\,\mathrm{Mpc}$ and employ $N_\mathrm{box}=1024^3$ simulation particles\footnote{The simulation particle mass is $m_\mathrm{sim}\simeq 1.01 \times 10^4\,h^{-1}\,\mathrm{M}_\odot$.}. The three models are evolved up to $z=14$, using a Plummer-equivalent gravitational softening length that is kept constant at $1/40$-th of the mean interparticle spacing. In the redshift ranges considered in our analysis, the $k$-modes with largest amplitudes just enter the non-linear regime by $z\approx 14$.

We identify collapsed structures using a friend-of-friends (FoF) algorithm with a linking length of $b=0.2$ times the mean interparticle spacing. We only consider FoF structures with more than $10^3$ particles, corresponding to a mass $M_\mathrm{DM}>M_\mathrm{threshold}\sim10^{7}\,h^{-1}\,\mathrm{M}_\odot$. Such structures are numerically well resolved and the simulations also resolve halo masses near the damping mass $M_d$ even in the coldest \wdm\ model. In addition, any {\em lower mass} objects have a virial temperature below $T\sim 10^4\,\mathrm{K}$ that is too low to enable cooling by \ion{H}{i} \cite{Barkana2001}. Gas in lower mass halos, if they were to form, would only cool through H$_2$ formation, but this channel is suppressed in \dmf\ models (see discussion in \cite{Boyarsky2019}). We sum all of the mass in collapsed structures to compute $f_{\rm coll}(z)$, the fraction of mass that is in collapsed objects. Note that the choice of the linking length to identify collapsed structures in simulations is somewhat uncertain (see {\em e.g.} \cite{Angulo2013} for the case of \wdm\ simulations). Understanding the full impact of different choices of $b$ on the fraction of collapsed objects at $z\sim 17$ is beyond the scope of this paper. However, in the next subsection, when acknowledging the possible uncertainties in our method for estimating the 21-cm absorption signal, we will briefly describe the expected overall effect of varying $b$ on $f_{\rm coll}$.

The evolution of $f_{\rm coll}$ in \dmf\ models can be understood by considering the evolution 
of the halo mass function, $n(M,z)$, discussed in the previous section in the context of the PS approach. In the PS model, $n(M,z)$ is a power-law at $M<\Mps(z)$, and exhibits an exponential cut-off at $M>\Mps(z)$. Here, $\Mps(z)$ is a characteristic mass which increases with time. At sufficiently high $z$, $M>\Mps(z)$, and $n(M,z)$ is exponentially small. As time increases, so does $\Mps$, until eventually $M\approx \Mps$, causing the abundance of structure of mass $M$, $n(M,z)$, to increase exponentially.  Eventually, $M\ll \Mps$, $n(M,z)$ remains on the power-law tail of the PS mass function and $n(M,z)$ evolves slowly. We can estimate the value of $\Mps(z)$
when objects start to form in our \dmf\ model, as follows. Setting $\Mps(z=0)\sim 10^{14}\,h^{-1}\,{\rm M}_\odot$ and taking the approximate growth rate
$M_h(z)=M_h(z=0)(1+z)^{0.24}\exp(-3z/4)$ from \cite{Correa2015}, yields $\Mps(z=20)\approx  10^8\,h^{-1}\,{\rm M}_\odot$, consistent with the numerical results of \cite{Reed2007}. Applying this reasoning to the special case of \dmf\ models, we infer that very little structure forms before $\Mps$, which is set by cosmology, becomes of order of the damping mass $M_d$, which is set by $\lambda_d$. As soon as these masses become comparable, structures will emerge and $f_{\rm coll}$ will increase exponentially. When $\Mps$ becomes much larger than $M_d$, the rate of increase of $f_{\rm coll}$ will decline. This expectation is borne-out by the simulations. In Figure~\ref{fig:1b}, we plot the total mass in collapsed objects in our $L_\mathrm{box}=5\,h^{-1}\,\mathrm{Mpc}$ simulations, for the three \dmf\ models. Coloured straight-lines are fits of the form
\begin{equation}
f_{\rm coll}(z) = f_{14}\exp(-\zeta(z-14))\,,    
\label{eq:fit1}
\end{equation}
to the simulation results, with parameters $f_{14}$ and $\zeta$ reported in Table~\ref{table:best_fit}.
The fits reproduce the simulation results well, and we use them to compute the evolution of $f_{\rm coll}$
in the \dmf\ models\footnote{We expect that the exponential accretion of the collapsed fraction (Eq.~(\ref{eq:fit1})) in \dmf\ will be replaced by a power-law evolution at later redshifts ($z<14$), similar to what found in~\cite{Correa2015} for \cdm.}.

We now have expressions for the fraction of mass in collapsed objects in which we assume that stars form, both in \cdm\ and in \dmf\ models. Next we describe how we use $f_{\rm coll}$ to describe the onset of star formation and the build-up of a \lya\ background. 

\subsection{Star formation and the build-up of a \lya\ photon background}
\label{sec:starformLyalpha}
\begin{figure}[t!]
\advance\leftskip-.15cm
\advance\rightskip-.5cm
\subfigure[][Lyman-$\alpha$ coupling, $x_\alpha$]
{\includegraphics[width=.52\textwidth]{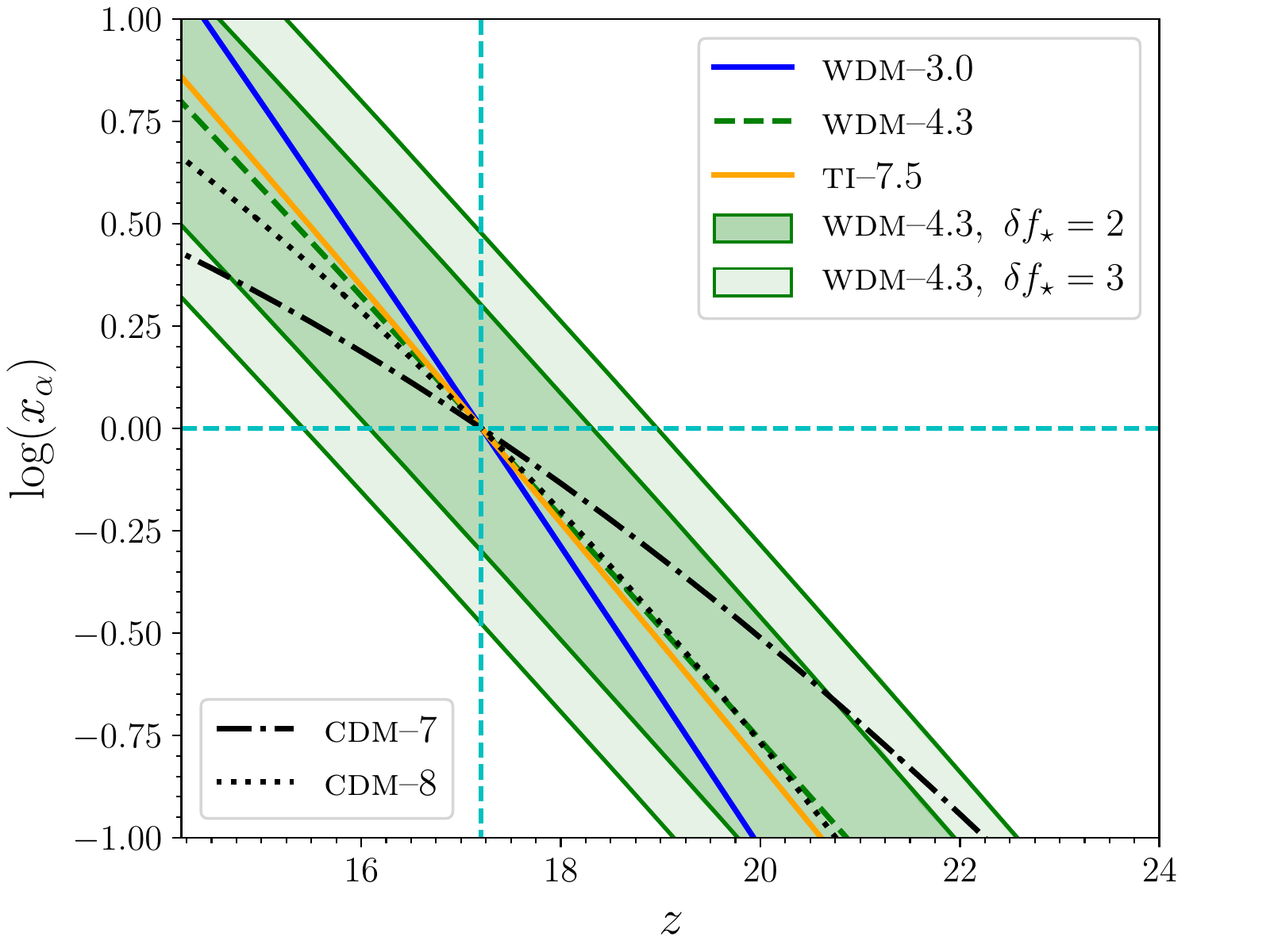}\label{fig:2a}}\hspace{-4.ex}
\subfigure[][Differential brightness temperature, $\delta T_\mathrm{b}$]{\includegraphics[width=.52\textwidth]{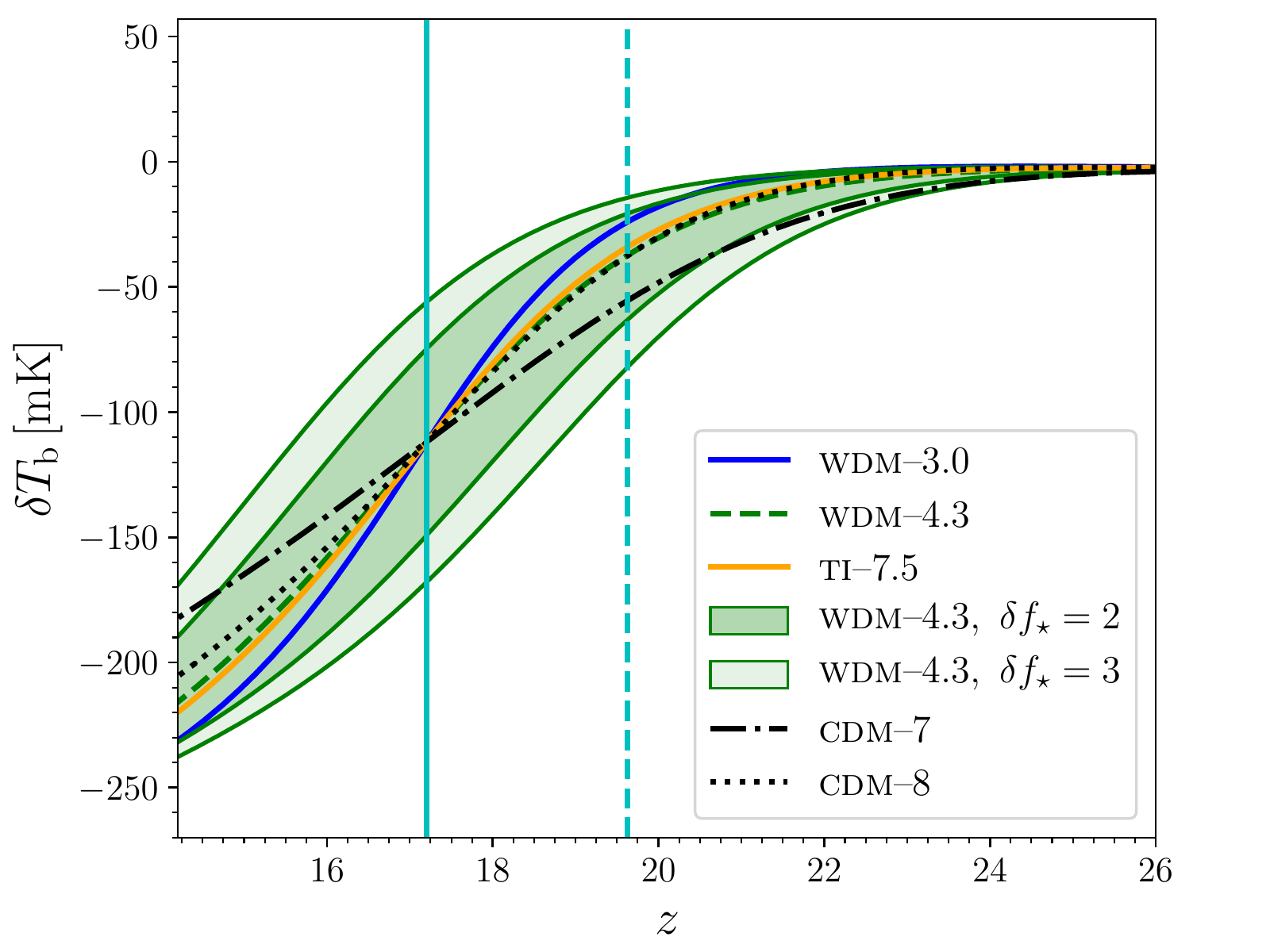}\label{fig:2b}}\\[-2. ex]
\caption{(a) Lyman-$\alpha$ coupling constant $x_\alpha$ for the five models considered here (different colour lines show different models as labelled), calculated imposing $x_\alpha=1$ at $z=17.2$, shown in the figure as intercept of the two cyan dashed lines. The colored bands show the values in the range $[x_\alpha/ \delta f_\star,x_\alpha\times \delta f_\star]$ for $\delta f_\star=2$ (dark green shaded area) and $\delta f_\star=3$ (light green shaded area). (b) Evolution of the differential brightness temperature $\delta T_\mathrm{b}$ as a function of the redshift $z$ for the different models, calculated from the $x_\alpha$ values in Figure~\ref{fig:2a}.  The cyan vertical solid line shows the redshift corresponding to the mean frequency of the {\sc edges} experiment, while the cyan vertical dashed line represents the redshift where the amplitude of the \edges\ signal is at half of its
maximum (these are the same as those shown in \cite{Schneider2018}). Note that we do not attempt to model the X-ray background heating of the hydrogen gas, which makes the signal disappear at late times. }
\label{fig:2}
\end{figure}

\begin{table}
\centering
\begin{tabular}{|c|c|c|c|c|c|}
\hline
Model&\wdm-3.0&\wdm-4.3&\ti-7.5&\cdm-7&\cdm-8\\
\hline\hline
$f_\star$     & $0.061$&$0.017$&$0.011$&$0.003$&$0.017$\\
\hline
\end{tabular}
\caption{Values of $f_\star$ obtained by imposing $x_\alpha = 1$ at $z=17.2$ as described in Section~\ref{sec:starformLyalpha}.}
\label{table:Nunknown_z17}
\end{table}

We characterize the star formation efficiency of collapsed structures by the parameter $f_\star$, which relates
the (co-moving) star formation rate density, $\dot{\rho}_\star$, to the rate at which structures collapse,
\begin{equation}
\dot{\rho}_\star(z) = f_\star \,\bar{\rho}^{0}_{b}\, \frac{df_\mathrm{coll}(z)}{dt}\,,
\label{eq:SFRD}
\end{equation}
where $\bar{\rho}^{0}_{b}$ is the present day baryon density. For a single galaxy in a halo, $f_\star$ sets the ratio of the stellar mass to halo mass,
\begin{equation}
f_\star = \frac{M_\star/\Omega_b}{M_h/\Omega_m}\,.
\end{equation}
This expression allows us to estimate a maximal value for $f_\star$. Ref.~\cite{Sharma2019} presents a model of feedback-regulated galaxy formation, in which the star formation rate of a galaxy is set by the balance between 
the energy lost by the deepening of the potential of its host dark matter halo due to cosmological accretion
and the energy injected by supernovae. The model predicts a ratio $M_\star/M_h\sim 10^{-3}$ for a halo of mass $M_h=10^8{\rm M}_\odot$ at $z\sim 17$, corresponding to $f_\star=0.5\%$. The cosmological hydrodynamical simulation presented by \cite{Sawala2015} give a similar median ratio of $M_\star/M_h$ in $M_h\sim 10^8{\rm M}_\odot$ halos, but with a relatively large scatter. Observations of satellites in the Milky Way also give a similar value for the stellar fraction at this halo mass (see \cite{Sawala2015}, their Fig.~4). We will consider a model to be viable provided $f_\star\in [0.1,2]\%$, that is within a factor of 4 larger or smaller than our best estimate.

We further assume that the co-moving UV-emissivity, $\epsilon_\nu(z)$, is proportional to the star formation rate,
\begin{equation}
\epsilon_\nu(z)   = \epsilon_\mathrm{b}(\nu)\,\frac{\dot{\rho}_\star(z)}{m_\mathrm{H}}\,,
\label{eq:emissivity_1}
\end{equation}
where $m_\mathrm{H}$ is the proton mass. Here, $\epsilon_\mathrm{b}(\nu)$ is the number of photons per unit of frequency emitted at frequency $\nu$ per baryon in stars. We assume that $\epsilon_\mathrm{b}(\nu)$ is constant over the interval $[\nu_\alpha, \nu_\mathrm{L}]$ (where $\nu_\mathrm{L}$ is the Lyman-limit frequency). We choose $\epsilon_\mathrm{b}(\nu)$ so that a given number, $N_\alpha$, of photons is produced per baryon in stars in the frequency interval $[\nu_\alpha, \nu_\mathrm{L}]$. If Pop.~II stars are the dominant sources of UV photons, then $N_\alpha \approx 9690$ \cite{Barkana2005}, using the {\sc starburst99} model by \cite{Leitherer1999}. Given $f_\mathrm{coll}(z)$, as computed in the previous section, these two equations yield $\epsilon_\nu(z)$, which allows the calculation of the specific mean intensity of \lya\ photons, $J_\alpha(z)$, using Eq.~(\ref{eq:Ja}). 

The calculation presented so far involves several uncertain parameters. The first parameter is our choice of linking length, $b$, used to identify collapsed structure in the \dmf\ simulations. Secondly, the star formation efficiency, $f_\star$, is not very well known. Previously we
argued that we expect that a reasonable model should have $f_\star\in [0.1,2]\%$ (in halos of mass $M_h\sim 10^8{\rm M}_\odot$), but in fact
$f_\star$ is likely to depend on halo mass (see {\em e.g.} \cite{Mirocha2017, Furlanetto2017, Sun2016} for detailed studies on the dependence of $f_\star$ on halo mass and redshift). Once stars start to form in a galaxy, supernovae associated with the end-stages of massive stars inject a large amount of energy into the galaxy, and this may strongly suppress further star formation, see {\em e.g.} \cite{Norman2018}. The importance of this feedback loop will depend on the nature of the galaxy -- in particular on the depth of its gravitational potential -- as well as on the nature of the stars. In addition, the minimum halo mass in which star formation will occur is not well known, as briefly discussed in the previous section. Finally, the function $\epsilon_\mathrm{b}(\nu)$ that relates $\dot\rho_\star$ to $\epsilon_\nu$ depends on the nature of the stars -- in particular on the initial stellar mass function -- which is not very well known.

To make progress, we proceed as follows. The exponential build-up of mass in \dmf\ models means that $\epsilon_\nu$ is mostly determined by the star formation efficiency of halos with mass around the damping mass, $M_d$, given our choice of models. The value of $f_{\rm coll}$ in such halos depends on the linking length $b$ - but a different choice of $b$ will simply result in a larger or smaller value of $f_{\rm coll}$ without affecting its evolution. As a consequence, the uncertainty in parameters - $b$, $f_\star$ and $\epsilon_\nu$ - will simply appear as an overall normalization constant in the value of $J_\alpha(z)$. Of course, the value of this normalization constant is of interest, yet our modelling is sufficiently uncertain that we cannot hope to calculate it with any real accuracy. Therefore, we instead {\em choose} $f_\star$ in each of our \dmf\ models such that the \lya\ coupling coefficient from Eq.~(\ref{eq:xa}) is unity at $z=17.2$ (the redshift corresponding to the mean frequency of the {\sc edges}  experimental absorption signal \cite{Bowman2018}), {\em i.e.} $x_\alpha(z=17.2)=1$. We choose this value because it gives a 21-cm absorption signal that is in relatively good agreement with the {\em timing} of the \edges\ detection. The required value for $f_\star$ for all \dmf\ models is specified in Table~\ref{table:Nunknown_z17}, given the evolution of $f_{\rm coll}$ and choice of linking length $b=0.2$ discussed in the previous section, and taking $N_\alpha=9690$. Given that we demand that a reasonable model should have $f_\star$ in the range of 0.1--2\%, the {\em timing} of the \edges\ signal seems to disfavour the WDM-3.0 model. In this model, structure formation is so much suppressed that the structures that do form need to be much more efficient in forming stars than what is currently thought reasonable.

Assuming that $f_\star$ is a constant is less well motivated for the \cdm\ case. Indeed, a relatively extended range of halo masses can in principle contribute to the build-up of $J_\alpha$, and it is quite unlikely that star formation is equally efficient in all these halos (see {\em e.g.} \cite{Mirocha2017, Furlanetto2017, Sun2016}). The values for $f_\star$ that yield $x_\alpha(z=17.2)=1$ for the two \cdm\ case with different choices for $M_{\rm min}$, are also given in Table~\ref{table:Nunknown_z17} (assuming our default value of $N_\alpha=9690$). Both models require reasonable values of $f_\star$.

The two key quantities $x_\alpha$ and $\delta T_\mathrm{b}$ describing the 21-cm absorption feature are computed using the Accelerated Reionization Era Simulations code ({\sc ares}) \cite{Mirocha2018,Mirocha2017, Mirocha2014, Mirocha2012}. We provide the code with the star formation rate density, Eq.~(\ref{eq:SFRD}). Note that we do not attempt to model the upturn of the absorption signal at lower redshifts, so we consider the background X-ray efficiency parameter and the ionizing photon efficiency parameter to be $f_\mathrm{X} = 0$ and $f_\mathrm{esc}\, N_\mathrm{ion} = 0$, respectively (see \cite{Mirocha2014} for a definition of $f_\mathrm{X}$, while $f_\mathrm{esc},\,N_\mathrm{ion}$ are introduced in Appendix~\ref{sec:reionisation}). 

The resulting evolution of the \lya\ coupling constant, $x_\alpha(z)$, is plotted in Fig.~\ref{fig:2a} for all five models; \cdm\ models are shown in {\em black}, \dmf\ models in {\em colour}. We note that $x_\alpha(z=17.2)=1$ for all models, by construction; {\em cyan dashed lines} are drawn at $z=17.2$ and $x_\alpha=1$, to guide the eye. The effect of increasing or decreasing $f_\star$ by a factor 2 and 3 for model \wdm-4.3, are shown by {\em dark} and {\em light green} shading, respectively. In all \dmf\ models,  $x_\alpha$ increases exponentially with time, reflecting the exponential increase in the collapsed fraction. Once scaled to have $x_\alpha(z=17.2)=1$, there is little difference between them. The \cdm-8 model, which has $M_{\rm min}=10^8h^{-1}~{\rm M}_\odot$ ({\em dotted black line}) looks very similar to the \dmf\ models. This is not surprising since we neglect any halos below $M_{\rm min}$ in the calculation of $f_{\rm coll}$ - effectively making the \cdm\ model behave like a \dmf\ model with $M_d\sim M_{\rm min}$ ($f_\mathrm{coll}$ in \cdm-8 is very similar to that in other \dmf\ models, especially \wdm-4.3, see Figure~\ref{fig:1b}). In all these models, $x_\alpha$ increases rapidly with time, from $\log(x_\alpha)=-0.5$ to $+0.5$ over a redshift extent $\Delta z\approx 3$. The build-up of $x_\alpha$ in the \cdm-7 model, which has $M_{\rm min}=10^7h^{-1}~{\rm M}_\odot$ ({\em dashed black line}), is considerably more extended in redshift, requiring $\Delta z\gtrsim 5$ for a ten-fold increase in $x_\alpha$. This is a direct result of lower-mass halos, whose abundance does not increase rapidly in time, contributing significantly to $J_\alpha$.

The corresponding evolution of the brightness temperature difference, $\delta T_b$, is shown in Fig.~\ref{fig:2b}, using the same colour/line style conventions.  The {\em cyan vertical solid line} at $z=17.2$ is the mean redshift of the \edges\ signal, while the {\em cyan vertical dashed line} represents the redshift where the  amplitude is at half of its maximum. They are drawn to roughly indicate the range of redshifts spanned by the absorption trough of the \edges\ signal in its downturn region.
Note that all the models in Fig.~\ref{fig:2b} predict $\delta T_\mathrm{b}\simeq -112~{\rm mK}$ at $z=17.2$, which is expected because we have scaled $f_\star$ to yield $x_\alpha=1$ at $z=17.2$ for all the models. As could be expected from the earlier discussion, the onset of 21-cm absorption is more rapid in the \dmf\ and \cdm-8 models, compared to the \cdm-7 model. However, deciding which, if any, of these look like the \edges\ detection is not
obvious. In particular, since we do not attempt to model the {\em decrease} of the absorption at lower $z$, thought to be caused by X-ray heating, we cannot compare the mean redshift of the simulated absorption line to the \edges\ data. Moreover, the absorption line is much stronger in the data than can be understood by simply coupling \Ts\ to \Tk\ through the WF-effect, as discussed in the Introduction.
According to \cite{Schneider2018}, $\delta T^\mathrm{min}_\mathrm{b}\in [-180,-100]\,\mathrm{mK}$ before the upturn caused by X-ray heating,
and models are allowed if the position of the minimum ($\delta T^\mathrm{min}_\mathrm{b}$) appears at $z\gtrsim 17.2$. Since all our models predict the same value of $\delta T_\mathrm{b}\in[-180,-100]\,\mathrm{mK}$ at $z=17.2$, they are all allowed based on the \cite{Schneider2018} criterion. 

However, from a comparison between our results in Figure~\ref{fig:2b} and the downturn of the {\sc edges} signal, we can conclude that the results for the \dmf\ and \cdm-8 models are overall in better agreement with the range of redshifts spanned by the observed absorption trough than those of the \cdm-7 model. Indeed, in the case of \cdm-7, the downturn of the brightness temperature starts at higher redshifts and its profile is considerably shallower. Note that the situation is even worse for \cdm\ had we allowed star formation with the same efficiency in halos with mass lower than $10^7h^{-1}~{\rm M}_\odot$, {\em e.g.} by invoking significant star formation through molecular cooling of gas. The impact of such \lq Pop.~III\rq\ star formation in \cdm\  is uncertain, because the build-up of a background of Lyman-Werner radiation by this mode of star formation leads to strong negative feedback, limiting the number of Pop~III stars that can form, see {\em e.g.} \cite{Machacek2001}.

We conclude noticing that pre-recombination differential streaming of baryons with respect to dark matter \cite{Tseliakhovich10} may affect the formation of the first stars and galaxies. However, these effects are thought to be small for halos of mass larger than $10^7h^{-1}~{\rm M}_\odot$, see {\em e.g.} \cite{Naoz2013, Richardson2013}, and hence would not affect our conclusions.

\section{Summary and discussion}
\label{sec:summary}

The 21-cm signal in the pre-reionization era can be used to constrain models with damped matter fluctuations on small scales, because these models introduce a scale below which there is a delay of structure formation with respect to \cdm{} models. Deriving constraints using 21-cm physics in a given cosmological model requires knowledge of several ingredients: ({\em i}) the evolution of the fraction of dark matter in collapsed structures that can form stars, $f_{\rm coll}$, ({\em ii}) the star formation efficiency of these halos, $f_{\star}$, and ({\em iii}) the rate at which stars produce \lya\ photons, for example quantified in terms of the number of \lya\ photons emitted per baryon in stars, $N_\alpha$. The signal shape also depends on the rate at which the gas is heated by X-rays, a process that we have not modelled. As stressed by \cite{Boyarsky2019}, all three of these ingredients are relatively poorly understood and introduce uncertainties into the calculation of the global 21-cm signal. In particular it is not even clear whether the emergence of the first star forming galaxies in \dmf\ models resembles that in \cdm: there are good reasons to suspect the existence of significant differences \cite{Gao2007, Hirano2017}. 

In \cite{Boyarsky2019}, the authors have shown that the constraints from \cite{Schneider2018} on the scale $M_d$ below which structure formation is depressed in \dmf\ models, can be loosened if a higher star formation efficiency parameter is chosen. However, in all these previous works, the value of $f_\star$ has been held fixed for all models. Here, we have taken a different approach, namely picking $f_\star$ for each model
such that it reproduces the timing of the 21-cm line, and contrasting the rate at which the 21-cm signal builds up. This aspect of the modelling is particularly relevant in terms of the {\em shape} of the signal. Our findings can be summarized as follows:
\begin{itemize}
    \item Warm dark matter models with thermal-equivalent particle mass $m_\mathrm{WDM}\sim 3\,\mathrm{keV}$ can produce an absorption signal in line with the timing of the \edges\ results but only if $f_\star\sim 6\%$. We argued that such a star formation efficiency is higher than values coming from predictions of current star formation models and observations of satellites in the Milky Way, disfavouring this model. The colder model with $m_\mathrm{WDM}> 4\,\mathrm{keV}$ requires $f_\star\lesssim 2\%$. Our model of thermal inflation, \ti-7.5, requires $f_\star\sim 1.1\%$. Given the uncertainties in the modelling, we argue that both these models are consistent with the timing of the \edges\ signal.
    \item A \cdm\ model in which star formation in halos below a mass of $M_{\rm min}=10^8h^{-1}~{\rm M}_\odot$ is assumed to be negligible, for example due to stellar feedback, requires $f_\star\sim 1.7\%$, and is almost indistinguishable from our \dmf\ models. From the point of view of the 21-cm physics, it will be hard to distinguish such a \cdm\ model from a \dmf\ model.
    \item Reducing the minimum mass for a halo to undergo star formation to $M_{\rm min}=10^7h^{-1}~{\rm M}_\odot$ in \cdm\ does lead
    to generic differences with \dmf\ models. In such a model, a larger fraction of \lya\ photons is produced by stars that form in low-mass halos. The number density of such halos increases only slowly with time around $z\sim 17$, and this results in a more extended onset of the 21-cm absorption signal. Moreover, the value of $f_\star$ required in such a model is only $f_\star=0.3\%$. If $f_\star$ were to remain constant, which is in fact unlikely, than such a low star formation efficiency results in reionization below $z\sim 5$ (the result is shown in the Appendix).
    Reducing the minimum mass to even lower values than $10^7h^{-1}~{\rm M}_\odot$ would strengthen the above conclusion.
    \item Taken at face value, none of our models results in an onset of the 21-cm signal that is as rapid as the observed \edges\ signal (c.f.~Figure~\ref{fig:2b}).
    However, the more that low-mass halos contribute to \lya\ photon production, the shallower the resulting onset. Therefore we find that \dmf\ models, if anything, are {\em preferred} by the \edges\ signal, rather than ruled out. \cdm\ models can still produce a rapid onset of 21-cm absorption, but only if the physics of star formation conspires with that of structure formation, to make the \cdm\ model mimic the \dmf\ model. An example is our \cdm-8\ model. 
    
\end{itemize}
Due to uncertainties in the physics of star formation, it is currently not possible to put strong constraints on \dmf\ models using the 21-cm absorption line. We expect that future studies will be able to provide answers to the above open questions, providing a better understanding of the star formation physics at the redshifts involved in the 21-cm global absorption profile.

We conclude commenting on how our results will change when considering larger or lower values of $M_\mathrm{min}$ than those considered here for \cdm.  A larger value for $M_{\rm min}$ than $10^8\,h^{-1}\,\mathrm{M}_\odot$ makes the 21-cm absorption signal set-in more suddenly, because of the more rapid increase in the number density of such halos around $z\sim 17$, compared to the case of \cdm-8. However, that also means that the star forming halos are rarer, and hence this requires a larger value of $f_\star$ in order to produce enough \lya\ photons by $z\sim 17$, even larger than the $\sim 1.7\%$ of the \cdm-8 model. Such high values of $f_\star$ are unlikely, both on theoretical grounds, and based on the observed low $M_\star/M_h$ fractions of present-day low-mass galaxies. Much {\em lower} values of $f_\star$ are possible when lowering $M_{\rm min}$ to values  $<10^7\,h^{-1}\,\mathrm{M}_\odot$. However, as commented above, this makes the onset of the 21-cm signal too shallow (much more shallower than that in \cdm-7), and is also more in tension with the onset of {\em reionization} than \cdm-7.

\acknowledgments

We thank Sownak Bose, Michael Buehlmann, Marius Cautun, Liang Gao, Oliver Hahn, John Regan and John Wise for valuable discussions. 
We also thank an anonymous referee for a very careful reading of the paper and their numerous comments that improved the manuscript. We are grateful to Alexey Boyarsky and Oleg Ruchayskiy for sharing their work \cite{Boyarsky2019} and for their comments on this manuscript. We thank John Helly for providing us with the software to identify collapsed regions,  Jordan Mirocha for his help with {\sc ares}, Mark Lovell, Arvind Kumar Mishra and Guochao Sun for their comments on the manuscript.  ML and BL are supported by the European Research Council via grant ERC-StG-716532-PUNCA. BL is additionally supported by STFC Consolidated Grants ST/P000541/1, ST/L00075X/1. SP is supported by the European Research Council under ERC Grant ``NuMass'' (FP7- IDEAS-ERC ERC-CG 617143) and acknowledges partial support from the Wolfson Foundation and the Royal
Society. SP, CMB and BL are also supported in part by the European Union's Horizon
2020 research and innovation program under the Marie Sklodowska-Curie grant agreements
No. 690575 (RISE InvisiblesPlus) and 674896 (ITN Elusives). This work used the DiRAC@Durham facility managed by the Institute for
Computational Cosmology on behalf of the STFC DiRAC HPC Facility
(www.dirac.ac.uk). The equipment was funded by BEIS capital funding
via STFC capital grants ST/K00042X/1, ST/P002293/1, ST/R002371/1 and
ST/S002502/1, Durham University and STFC operations grant
ST/R000832/1. DiRAC is part of the National e-Infrastructure.

\appendix
\section{Reionization}
\label{sec:reionisation}

\begin{figure}[t!]
\centering
{\includegraphics[width=.6\textwidth]{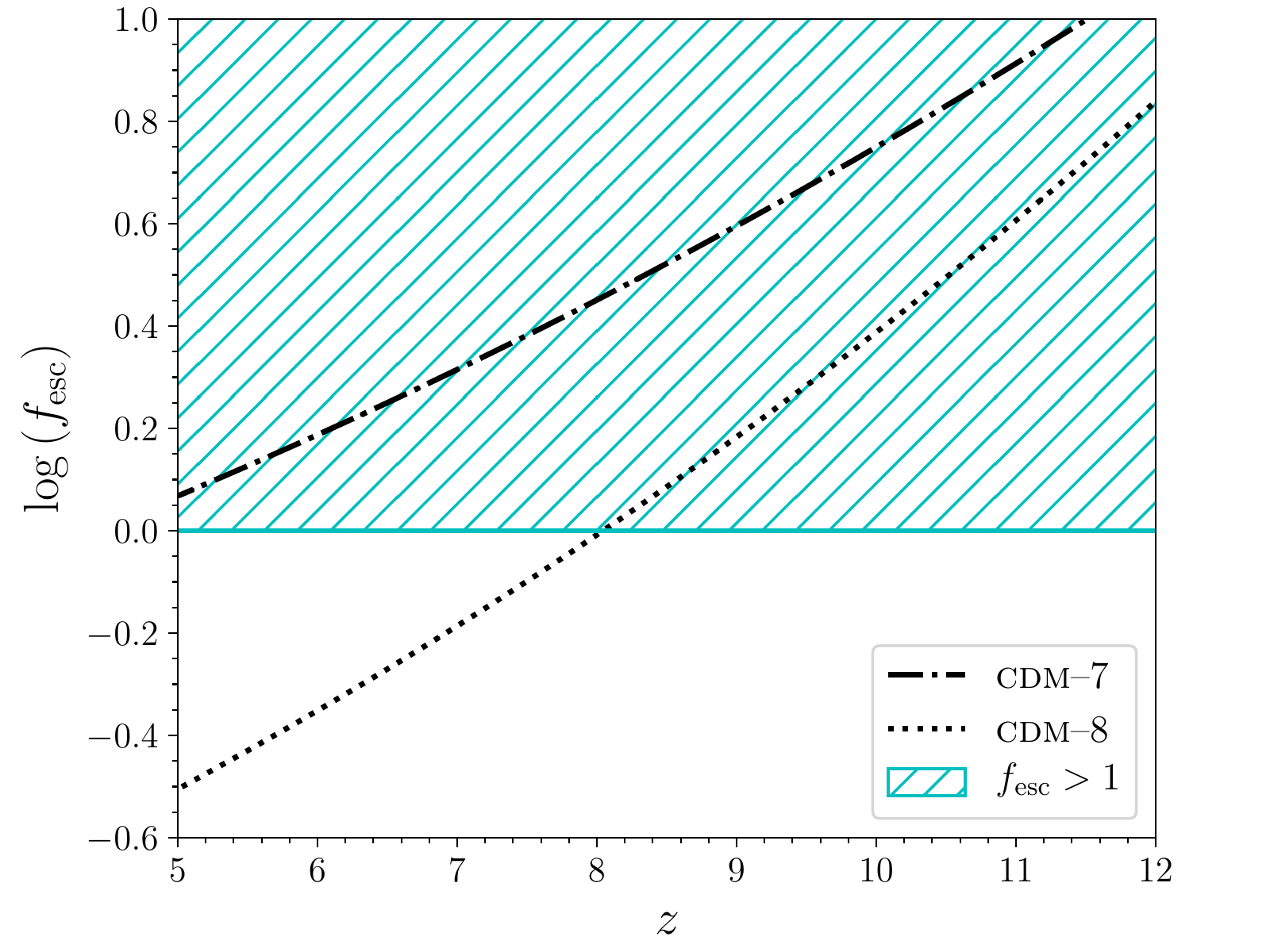}}\\[-2. ex]
\caption{Escape fraction of ionizing photons, $f_\mathrm{esc}$, needed to achieve $x_\mathrm{ion}=2$ at a given redshift, for the two \cdm\ models \cdm-7 and \cdm-8 (as labelled). The shaded area shows the region where $f_\mathrm{esc}>1$. Since  $f_\mathrm{esc}\leq 1$ by definition, this region is not allowed.}
\label{fig:ionisation_photons}
\end{figure}

Here, we address the reionization process in \cdm-7 and \cdm-8. To do so, we estimate the cumulative number density of ionizing photons with energy between $\left[13.6, 24.6\right]\,\mathrm{eV}$ produced at a given redshift as,
\begin{equation}
n^\mathrm{ion}_\gamma(z) = \int^\infty_z dz^\prime \,\frac{dn^\mathrm{ion}_\gamma (z^\prime)}{dz^\prime},    
\end{equation}
where ${dn^\mathrm{ion}_\gamma}/{dz}$ is the  number density of ionizing photons produced in the time interval corresponding to $dz$ (the number of ionizing photons is calculated using {\sc ares} as in \cite{Mirocha2015,Mirocha2014}). ${dn^\mathrm{ion}_\gamma}/{dz}$ can be given in terms of the ionization rate \cite{Mirocha2015},
\begin{equation}
\Gamma_\mathrm{HI}  = N_\mathrm{ion}\,f_\mathrm{esc}\, \dot{\rho}_\star   
\end{equation}
which depends on the star formation rate density, $\dot{\rho}_\star$ (note that $\dot{\rho}_\star$ depends on $f_\star$, see Eq.~(\ref{eq:SFRD})) and on the fraction of ionizing photons (per stellar baryon) that can escape from their host galaxies, $f_\mathrm{esc}\, N_\mathrm{ion}$, where $N_\mathrm{ion}$ is the number of ionizing photons emitted per stellar baryon. Here, we consider $N_\mathrm{ion} = 4000$ \cite{Barkana2001}. We approximately estimate the redshift of reionization, $z_\mathrm{ion}$, as the redshift by which, cumulatively, two ionizing photons per baryon were emitted, $x_\mathrm{ion}= 2$ ($x_\mathrm{ion}\equiv n^\mathrm{ion}_\gamma/n_{b}$ and $n_{b}$ is the number density of baryons). Since $x_\mathrm{ion}$ depends on $f_\star \, f_\mathrm{esc}$, taking $f_\star$ from Table~\ref{table:Nunknown_z17}, we can estimate for each model the escape fraction needed to achieve $x_\mathrm{ion} = 2$. This result is shown in  Figure~\ref{fig:ionisation_photons}. 
Since $f_\mathrm{esc}\leq 1$ by definition, from Figure~\ref{fig:ionisation_photons} we conclude that a value of $f_\star = 0.003$ (that produces an absorption trough in line with the timing of the {\sc edges} signal, see Section~\ref{sec:starformLyalpha}) for \cdm-7 cannot ensure reionization at redshifts $z>5$. On the other hand, $f_\mathrm{esc}\leq 1$ at $z\leq 8$ in the case of \cdm-8, in better agreement with the evidence of reionization at $z\approx 7$.

\end{document}